\documentclass[pre,superscriptaddress,amsfonts,showpacs]{revtex4}
\usepackage{epsf}

\newcommand{\bd     }{\begin{displaymath}}
\newcommand{\ed     }{\end{displaymath}}

\newcommand{\s      }{\sigma}

\newcommand{\mc     }{\mathcal}
\newcommand{\bra    }{\langle}
\newcommand{\ket    }{\rangle}

\newcommand{\atanh  }{{\rm{ath}}}

\newcommand{\bsigma }{\mbox{\boldmath$\sigma$}}

\newcommand{\bone   }{\mbox{\boldmath$1$}}

\newcommand{\btau   }{\mbox{\boldmath$\tau$}}

\newcommand{\br     }{\mbox{\boldmath$r$}}

\newcommand{\bxi    }{\mbox{\boldmath$\xi$}}
\newcommand{\bzeta  }{\mbox{\boldmath$\zeta$}}

\newcommand{\bI     }{\mbox{\boldmath$I$}}
\newcommand{\mcA    }{\mathcal{A}}
\newcommand{\mcG    }{\mathcal{G}}
\newcommand{\LK     }{\Lambda_K}
\newcommand{\OL     }{\Omega_L}
\newcommand{\LKP    }{\Lambda_{K^\prime}}
\newcommand{\OLP    }{\Omega_{L^\prime}}
\newcommand{\CB     }{\overline{C}}
\newcommand{\LB     }{\overline{L}}

\begin{document}

\title{\bf
Analysis of common attacks in LDPCC-based public-key cryptosystems}
\author{{\bf N.S.~Skantzos$^{\dagger *}$, D.~Saad$^{\dagger}$
and Y.~Kabashima$^{\ddagger}$}\\
$^{\dagger}$ Neural Computing Research Group, Aston University, B4 7ET, UK\\
$^{*}$ Institut for Theoretical Physics, Celestijnenlaan 200D,
KULeuven, Leuven, B-3001 Belgium \\
$^{\ddagger}$ Dept.\@ of Computational Intelligence \& Systems Science,\\
Tokyo Institut of Technology, Yokohama  2268502, Japan\\
email: \texttt{skantzon@aston.ac.uk}, \texttt{saadd@aston.ac.uk}
and \texttt{kaba@dis.titech.ac.jp}}

\begin{abstract}
\noindent We analyze the security and reliability of a recently
proposed class of public-key cryptosystems against attacks by
unauthorized parties who have acquired partial knowledge of one or
more of the private key components and/or of the plaintext. Phase
diagrams are presented, showing critical partial knowledge levels
required for unauthorized decryption.
\end{abstract}

\maketitle

%--------------------------------------------------------------------%
\section{Introduction}
%--------------------------------------------------------------------%
%
An important aspect in many modern communication systems is the
ability to exclude unauthorized parties from gaining access to
confidential material. Although cryptosystems in general have an
extensive history, until fairly recently they have been based on
simple variations of the same theme: information security among
authorized parties relies on sharing a secret key which is to
be used for encryption and decryption of transmitted messages.
While in this way confidentiality of the sent message may be
secured, such systems suffer from the (obvious) drawback of
non-secure key distribution.

In 1978 Rivest, Shamir and Adleman first devised a way to resolve
this problem which led to the celebrated RSA \emph{public-key}
cryptosystem~\cite{RSA} (for historical accuracy, a similar system
has been suggested years earlier in the British GCHQ but was kept
secret). The idea behind public key cryptosystems is to
differentiate between the encryption- and decryption-keys; private
key(s) are assigned to authorized users, for decryption purposes,
while transmitting parties only need to know the matching
encryption (public) key~\cite{DH}. The two keys are related by a
function which generates the encryption mechanism from the
decryption key with low computational costs, while the opposite
operation (evaluating the decryption key from the encryption
mechanism) is computationally infeasible. Such functions are
called `one-way' or trap-door functions; the RSA algorithm for
instance, is based on the intractability of factorizing large
integers generated by taking the product of two large prime
numbers.

The proliferation of digital communication in the last few decades
has brought in a demand for secure communication leading to the
invention of several other public-key cryptosystems, most notable
of which are the El-Gammal cryptosystem (based on the Discrete
Logarithm problem), systems based on elliptic curves and the
McEliece cryptosystem (based on linear error-correcting
codes)~\cite{stinson}. A common denominator of all public-key
algorithms is the high computational complexity  of the task
facing the unauthorized user; this is typically related to hard
computational problems that cannot be solved in practical time
scales.

A new public-key cryptosystem based on a diluted Ising spin-glass
system has been recently proposed in~\cite{KMS}. The suggested
cryptosystem is similar in spirit to that of McEliece and relies
on exploiting physical properties of the MacKay-Neal (MN)
low-density parity-check (LDPC) error-correcting codes. In
particular, in the context of MN codes it has been
shown~\cite{KMS,mackay,kabashima} that for certain parameter
values successful decoding is highly likely, while for others
(particularly when the number of parity-checks per bit and the
number of bits per check tend to infinity) the `perfect' solution,
describing full retrieval of the sent message, admits only a very
narrow basin of attraction; iterative algorithmic solutions lead
in this case, almost certainly, to a decryption failure. One can
use these properties to devise an LDPC based
cryptosystem~\cite{KMS}. The narrow basin of attraction ensures
that a random initialization of the decryption equations will fail
to converge to the plaintext solution while the naive approach of
trying all possible initializations is clearly doomed for a
sufficiently large plaintext size. The `one-way' function relies
on the hard computational task of decomposing a dense matrix (the
public key) into a combination of sparse and dense matrices
(private keys)~\cite{garey_johnson}.

In this paper we examine the suggested cryptosystem from an
adversary's viewpoint. We consider an unauthorized party that has
acquired partial or full knowledge of one or more of the private
keys, and/or of the message, and we evaluate the critical
knowledge levels required for unauthorized decryption. In
addition, we examine the decryption reliability by authorized
users due to the probabilistic nature of the cryptosystem.

The paper is organized as follows: In the following section we give an
outline of the suggested cryptosystem. In section~\ref{sec:attack} we
formulate unauthorized-decryption scenarios with partial knowledge
based on a statistical mechanical framework. In
section~\ref{sec:analysis} we derive the observable quantity that
measures decryption success of the unauthorized user as a function of
the attack parameters and in section~\ref{sec:results} we examine
various cases and present numerical results as well as the related
phase diagrams. In sections~\ref{sec:BOA} and~\ref{sec:reliability} we
briefly study the basin of attraction of the ferromagnetic solution,
and the reliability of the decryption mechanism (for authorized
users), respectively. The implication of the analysis are discussed in
section~\ref{sec:conclusion}.

%--------------------------------------------------------------------%
\section{Description of the Cryptosystem}
%--------------------------------------------------------------------%
%
The cryptosystem suggested in~\cite{KMS} is based on the framework
of MN error-correcting codes~\cite{mackay}. An outline of the
encryption/decryption process is as follows.

A plaintext represented by $\bxi\in\{0,1\}^N$ is encrypted to the
ciphertext $\br\in\{0,1\}^M$ (with $M>N$) using a predetermined
generator matrix $G\in\{0,1\}$ and a corrupting vector
$\bzeta\in\{0,1\}^M$ with $P(\zeta_i)=p \ \delta_{\zeta_i,1}+(1-p)
\ \delta_{\zeta_i,0}$ for each component $1 \le i \le M$; the
Kronecker tensor $\delta_{ab}$ returns 1 when the arguments are
equal ($a=b$) and zero otherwise. The generated ciphertext is of
the form:
\begin{equation}
\br=G\bxi+\bzeta
\hspace{10mm}
{\rm (mod\ 2)}
\end{equation}
The  $(M\times N)$ matrix $G$ together with the corruption rate
$p\in[0,1]$ constitute the \emph{public} key.

The encryption matrix $G$ is constructed by choosing a dense
matrix $D$ (of dimensionality $M\times M$) and two
randomly-selected sparse matrices $A$ (of dimensionality $M\times
N$) and $B$ (of dimensionality $M\times M$) through $G=B^{-1}A D$
(mod 2). The matrices $A$ and $B$ are characterized by $K$ and $L$
non-zero elements per row and $C$ and $L$ non-zero elements per
column respectively. The resulting dense matrix $G$ is {\em
modeled} as being characterized by $K^\prime$ and $C^\prime$
non-zero elements per row and per column respectively with
$K^\prime,C^\prime\to\infty$ (while $K^\prime/C^\prime=N/M$ is
finite). In fact, the dense matrix $G$ is of an irregular form due
to the inverse of the sparse matrix $B$ as well as the product
taken with the dense matrix $D$; we will model the matrix $G$ by a
regular dense matrix to simplify the analysis. The parameters
$K,C$ and $L$ define a particular cryptosystem while the matrices
$A$, $B$ and $D$ constitute the \emph{private} key.

The authorized user may obtain the plaintext from the received
ciphertext $\br$ by taking the (mod 2) product
$B\br=A\bxi+B\bzeta$. Finding a set of solutions $\bsigma$ and
$\btau$ such that the equation
\begin{equation}
A\bsigma+B\btau=A\bxi+B\bzeta
\hspace{10mm}
{\rm (mod\ 2)}
\label{eq:MN}
\end{equation}
is true will lead to candidate solutions of the decryption problem
(of which the most probable one will be detected according to a
further selection criterion). For particular choices of $K$ and
$L$, solving the above equation can be achieved via iterative
methods which have common roots in both graphical models and
physics of disordered systems such as Belief
Propagation~\cite{mackay} Belief Revision~\cite{weiss} and more
recently Survey Propagation~\cite{MPZ}; where state probabilities
for the decrypted message bits $P(\bsigma, \btau|\br)$ are
calculated by solving iteratively a set of coupled equations,
describing conditional probabilities of the ciphertext bits given
the plaintext and vice versa. This problem is identical to the
decoding problem of a regular MN error-correcting code; for the
explicit iterative decoding equations see
equations~(\ref{eq:horizontal}-\ref{eq:vertical}) as well
as~\cite{mackay,MKSV}.

The unauthorized user, on the other hand, faces the task of
finding the most probable solutions to the equation
\begin{equation}
G\bxi+\bzeta=G\bsigma+\btau \hspace{10mm} {\rm (mod\ 2)} \ .
\label{eq:sourlas}
\end{equation}
The above decryption equation is effectively identical to the decoding
problem of Sourlas error-correcting codes~\cite{sourlas}, with the
public matrix $G$ being dense. Most notably, in the context of Sourlas
codes, finding solutions to (\ref{eq:sourlas}) is strongly dependent
on initial conditions: for all initial conditions other than the
plaintext itself, the iterative equations of Belief Propagation will
fail to converge to the plaintext
solution~\cite{KMS,mackay,kabashima,kanter} such that obtaining the
correct solution for (\ref{eq:sourlas}) without knowledge of the
private key will become infeasible. Obtaining the private keys by
decomposing $G$ into $A$, $B$ and $D$ is known to be a hard
computational problem even if the values of $K$, $C$ and $L$ are
known~\cite{garey_johnson}.

We would like to point to the fact that there may exist more than one
triplet of matrices $\{A,B,D\}$ such that $G=B^{-1}AD$. with $D$ being
a dense matrix, finding a set of matrices $A^\prime$, $B^\prime$ and
$D^\prime$ such that their combination produces
$G=(B^\prime)^{-1}A^\prime D^\prime$ requires an exponentially
diverging number of operations, with respect to the system size,
making the decomposition computationally infeasible. For $D=\bone$ (as
was the original formulation in~\cite{KMS}) finding a pair of sparse
matrices $A^\prime$ and $B^\prime$ such that
$G=(B^\prime)^{-1}A^\prime$ requires only a number of operations that
is polynomial in $N$, and the cryptosystem is therefore not secure.

Other advantages and drawbacks of the new cryptosystem appear
in~\cite{KMS}.

%--------------------------------------------------------------------%
\section{Formulation of the Attack}
         \label{sec:attack}
%--------------------------------------------------------------------%
%
An essential ingredient of any cryptosystem is a certain level of
robustness against attacks. The robustness of the current cryptosystem
against attacks with no additional secret information has already been
reported in~\cite{KMS}. In this section we study the vulnerability of
the new cryptosystem to various attacks, characterized by partial
knowledge of the secret keys and/or the plaintext itself; the
additional information manifests itself in a set of decryption
equations similar to (\ref{eq:MN}) in which partial information of the
secret keys (and plaintext) is used in conjunction with the publicly
available information of (\ref{eq:sourlas}).
\begin{center}
\begin{figure}[b]
\setlength{\unitlength}{1.1mm}
\begin{picture}(120,55)
\put( 30,  0){\epsfysize=45\unitlength\epsfbox{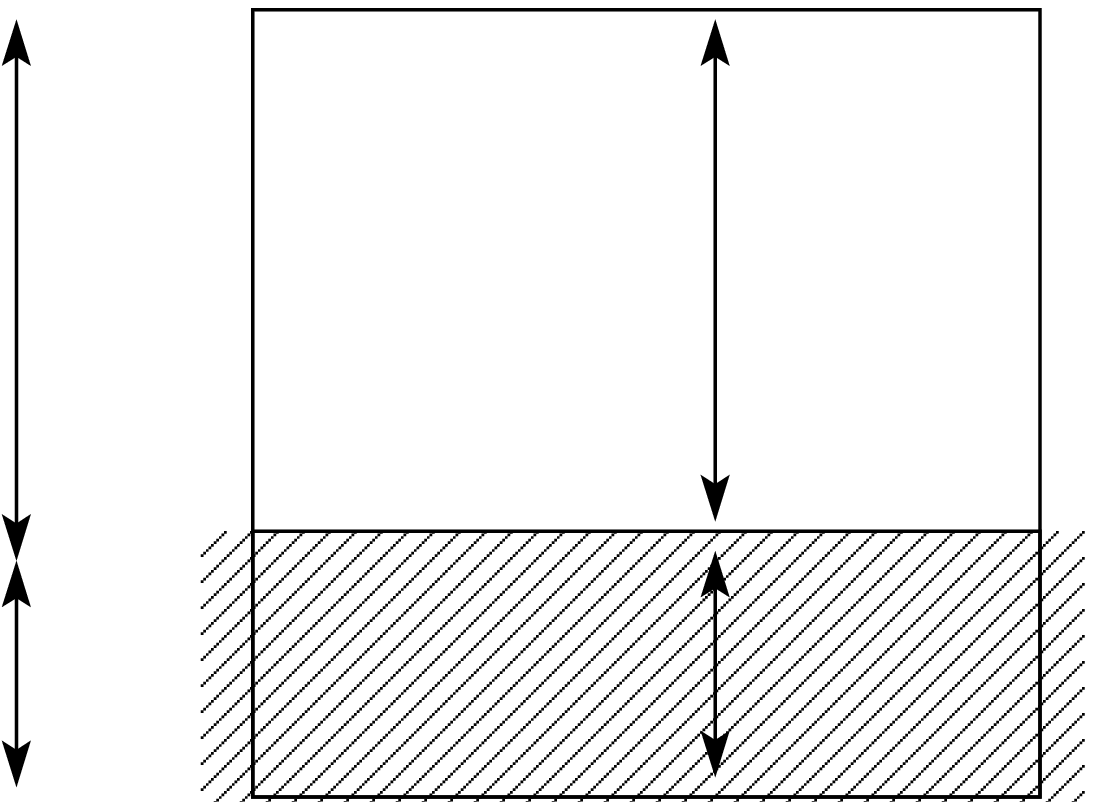}}
\put(12,27){$(1-\gamma)M$} \put(20,7) {$\gamma M$}
\put(58,27){$L-\tilde{L}_j$} \put(63,7) {$\tilde{L}_j$}
\end{picture}
\caption{ The matrix $B$ of dimensionality $M\times M$ used as a
private key in decryption. The scenario we consider here is that
unauthorized users have acquired knowledge of $\gamma M$ rows of
the matrix. The $(\gamma M\times M)$ block may have
$\tilde{L}_j=0,\ldots,L$ non-zero elements per column for all $j$.
} \label{fig:matrix}
\end{figure}
\end{center}
The cumulative information provided by the different sets of equations
will potentially allow for a successful decryption. To this extent,
knowledge of the matrix $B$ is of utmost importance since obtaining
partial knowledge of the syndrome vector and equation (\ref{eq:MN}) is
only accessible through decryption using the matrix $B$. Let us consider
that an unauthorized user has acquired knowledge of a number of rows
$\gamma_A M$, $\gamma_B M$ and $\gamma_D M$ of the secret matrices
$A$, $B$ and $D$ (with $\gamma_\star\in[0,1]$). Relation (\ref{eq:MN})
then provides $\gamma M\equiv{\rm min}\{\gamma_A,\gamma_B,\gamma_D\}
M$ decryption equations (\ref{eq:sparse-part}) based on sparse
matrices. To analyze the attack we will thus from now on assume that a
block $(\gamma M\times M)$ of all matrices is known to the
unauthorized user with $\gamma\in[0,1]$. In this case, the products
$\sum_{j=1}^MB_{ij}r_j$ for $i=1,\ldots,\gamma M$ can be taken and the
unauthorized user will arrive at the following decryption problem:
\begin{eqnarray}
{\rm private:}
&
(\hat{A}\bsigma)_i+(\hat{B}\btau)_i=(\hat{A}\bxi)_i+(\hat{B}\bzeta)_i
&
{\rm for\ rows}\ i=1,\ldots,\gamma M
\label{eq:sparse-part}
\\
{\rm public:}
&
(G\bsigma)_i+(\bI\btau)_i=(G\bxi)_i+(\bI\bzeta)_i
&
{\rm for\ rows}\ i=1,\ldots,M
\label{eq:dense-part}
\end{eqnarray}
where we absorbed the matrix $D$ using $\bsigma\to D\bsigma$ and
$\bxi\to D \bxi$; in practice, after decryption, one will have to
use of the inverted matrix $D^{-1}$ to obtain the original
plaintext. All solutions $\bsigma$ and $\btau$ will have to
simultaneously satisfy (\ref{eq:sparse-part}) and
(\ref{eq:dense-part}). The matrices $\hat{A}$ and $\hat{B}$ will
be described by $K$ and $L$ non-zero elements per row. The average
number of known non-zero elements per column in $\hat{A}$ and
$\hat{B}$ will be denoted $\CB$ and $\LB$, respectively. Since
$\gamma$ is the probability of selecting a non-zero element in the
known part of the private key it follows that $\CB=\gamma C$
and $\LB=\gamma L$. For all columns $j=1,\ldots,M$ we will denote
the number of non-zero elements in $\hat{A}$ and $\hat{B}$ by the
random variables $\tilde{C}_j(=\sum_{i=1}^{\gamma M}\hat{A}_{ij})$
and $\tilde{L}_j(=\sum_{j=1}^{\gamma M}\hat{B}_{ij})$ which are
described by the distributions:
\begin{eqnarray}
P(\tilde{C}_j;C)= & \left(\!\begin{array}{c} C \\ \tilde{C}_j
\end{array}\!\right)\ \gamma^{\tilde{C}_j}\
(1-\gamma)^{C-\tilde{C}_j} & \hspace{3mm} \tilde{C}_j=0,\ldots,C
\label{eq:prob_C}
\\
P(\tilde{L}_j;L)= & \left(\!\begin{array}{c} L \\ \tilde{L}_j
\end{array}\!\right)\ \gamma^{\tilde{L}_j}\
(1-\gamma)^{L-\tilde{L}_j} & \hspace{3mm} \tilde{L}_j=0,\ldots,L
\label{eq:prob_L}
\end{eqnarray}

To facilitate the statistical mechanical description we will now
replace the field $\{0,1;+{\rm (mod\ 2)}\}$ by the more familiar
Ising spin representation~\cite{sourlas} $\{-1,1;\times\}$. Equations
(\ref{eq:sparse-part}) and (\ref{eq:dense-part}) will also be
modified: From the matrices $\hat{A},\hat{B}$ and $G,\bI$ we
construct the binary tensors $\mcA=\{\mcA_{\bra i_1\cdots
i_K;j_1\cdots j_L\ket};1\leq i_1<\cdots<i_K\leq N, 1\leq
j_1<\cdots<j_L\leq M\}$ and $\mcG=\{\mcG_{\bra i_1\cdots
i_{K^\prime};j\ket}; 1\leq i_1<\cdots<i_{K^\prime}\leq N, 1\leq
j\leq M\}$. The elements of these tensors are $\mcA_{\bra
i_1\ldots i_K;j_1\ldots j_L\ket}=1$ if $\hat{A}$ and $\hat{B}$
have respectively a row in which the elements $\{i_1,\ldots,i_K\}$
and $\{j_1,\ldots,j_L\}$ are all 1 and 0 otherwise. Similarly,
$\mcG_{\bra i_1\ldots i_{K^\prime};j\ket}=1$ if $G$ and $\bI$ have
respectively a row in which the elements
$\{i_1,\cdots,i_{K^\prime}\}$ and $\{j\}$ are all 1 and 0
otherwise. The notation we used to indicate tensor elements, $\bra
i_1\ldots i_K\ket$, denotes that the sites $i_1,\ldots,i_K$ are
ordered and different.

The fact that the number of non-zero elements per column
in $\hat{A},\hat{B}$ and $G,\bI$, respectively, are $\tilde{C}_i, \tilde{L}_i$
and $C^\prime, 1$, for all columns,
will be imposed by the constraints:
\begin{eqnarray}
\sum_{i_2\cdots i_K;j_1\cdots j_L}
\mcA_{\bra i_1\cdots i_K;j_1\cdots j_L\ket}=\tilde{C}_{i_1}
&&
\forall i_1=1,\cdots,M
\label{eq:mcA1}
\\
\sum_{i_1\cdots i_K;j_2\cdots j_L}
\mcA_{\bra i_1\cdots i_K;j_1\cdots j_L\ket}=\tilde{L}_{j_1}
&&
\forall j_1=1,\cdots,M
\\
\sum_{i_2\cdots i_{K^\prime};j}
\mcG_{\bra i_1\cdots i_{K^\prime};j\ket}=C^\prime
&&
\forall i_1=1,\cdots, M
\\
\sum_{i_1\cdots i_{K^\prime}} \mcG_{\bra i_1\cdots
i_{K^\prime};j\ket}=1 && \forall j=1,\cdots,M \label{eq:mcG2}
\end{eqnarray}
To compress notation in what follows we will denote the set of
indices involved in the tensors $\mcA$ and $\mcG$ by $\LK=\bra
i_1\cdots i_K\ket$ and $\OL=\bra j_1\cdots j_L\ket$.

For the system described in
(\ref{eq:sparse-part}-\ref{eq:dense-part}) the microscopic state
probability $P(\bsigma,\btau)$ can be written as
\begin{equation}
P(\bsigma,\btau| \bxi,\bzeta, \mcA, \mcG)=\frac1Z\
\left[\Delta(\bsigma,\btau; \bxi,\bzeta, \mcA)\
      \Delta(\bsigma,\btau;\bxi,\bzeta,\mcG)\
      \Phi(\bsigma; \bxi)\
      \Phi(\btau; \bzeta)\right]\ e^{-\beta H(\bsigma,\btau)}
\label{eq:state_prob}
\end{equation}
(notice that the dependence on $\bxi,\bzeta$ is not explicit, but
through the received vector $\br$) where $Z$ is the partition
function and $H(\bsigma,\btau)$ the energy:
\begin{equation}
H(\bsigma,\btau)=-F_{\sigma}\sum_{i=1}^N\s_i-F_{\tau}\sum_{j=1}^M\tau_j
\label{eq:hamiltonian}
\end{equation}
with $F_{\sigma}=\frac12\log\frac{1-p_{\sigma}}{p_{\sigma}}$ and
$F_{\tau}=\frac12\log\frac{1-p_{\tau}}{p_{\tau}}$. The fields
$F_{\sigma}$ and $F_{\tau}$ represent prior knowledge of the
statistics from which the plaintext and the corrupting vector are
drawn, such that
\begin{equation}
P(\xi_i)=(1-p_{\sigma})\delta_{\xi_i,1}+p_{\sigma}\delta_{\xi_i,-1}
\hspace{10mm} p_{\sigma}\in[0,1] \label{eq:prob_xi}
\end{equation}
\begin{equation}
P(\zeta_j)=(1-p_{\tau})\delta_{\zeta_j,1}+p_{\tau}\delta_{\zeta_j,-1}
\hspace{10mm} p_{\tau}\in[0,1] \label{eq:prob_zeta}
\end{equation}
The indicator functions  $\Delta(\bsigma,\btau;\bxi,\bzeta,\mcA)$ and
$\Delta(\bsigma,\btau;\bxi,\bzeta,\mcG)$ restrict the  space of solutions
$\bsigma\in\{-1,1\}^N$ and $\btau\in\{-1,1\}^M$ to those that obey
equations (\ref{eq:sparse-part}) and (\ref{eq:dense-part}):
\begin{eqnarray}
\Delta(\bsigma,\btau; \bxi,\bzeta,\mcA)
&=&
\prod_{\LK\OL}\left[1+\frac12 \mcA_{\LK\OL}(\prod_{i\in\LK}\s_i\xi_i
\prod_{j\in\OL}\tau_j\zeta_j-1)\right]
\label{eq:Delta_A}
\\
\Delta(\bsigma,\btau; \bxi,\bzeta, \mcG)
&=&
\prod_{\LKP\OLP}\left[1+\frac12 \mcG_{\LKP\OLP}(\prod_{i\in\LKP}\s_i\xi_i
\prod_{j\in\OLP}\tau_j\zeta_j-1)\right]
\label{eq:Delta_G}
\end{eqnarray}
and finally the terms $\Phi(\cdots)\in\{0,1\}$ correspond to
\begin{eqnarray}
\Phi(\bsigma; \bxi)
&=&
\prod_{i=1}^N\left[(1-c_i)+c_i\delta_{\sigma_i,\xi_i}\right]
\\
\Phi(\btau; \bzeta)&=&
\prod_{i=1}^M\left[(1-d_i)+d_i\delta_{\tau_i,\zeta_i}\right]
\end{eqnarray}
where the quenched variables $c_i,d_j\in\{0,1\}$ model prior knowledge
of bits of the plaintext and the corrupting vector such that
if for some $i$ the plaintext bit $\xi_i$ is known
then the thermal variable $\s_i$ takes the quenched plaintext value
(and similarly for the corruption vector $\zeta_j$ and $\tau_j$).
For the distribution of $c_i$ and $d_j$ we will consider
\begin{equation}
P(c_i)=w_{\sigma}\,\delta_{c_i,1}+(1-w_{\sigma})\,\delta_{c_i,0}
\hspace{10mm} w_{\sigma}\in[0,1] \label{eq:prob_wp}
\end{equation}
\begin{equation}
P(d_j)=w_{\tau}\,\delta_{d_j,1}+(1-w_{\tau})\,\delta_{d_j,0}
\hspace{10mm} w_{\tau}\in[0,1] \label{eq:prob_wc}
\end{equation}
The system described by (\ref{eq:state_prob}) represents a set of
variables interacting via multi-spin ferromagnetic couplings of
finite connectivity, represented by a combination of matrices, in
the presence of the random fields $\xi_iF_{\sigma}$ and
$\zeta_jF_{\tau}$. At $\beta=1$ (which corresponds to the
Nishimori temperature \cite{nishimori}) we will evaluate the free
energy per plaintext bit
\begin{equation}
f=-\lim_{N\to\infty}\frac{1}{\beta N}\left\bra \log
Z\right\ket_{\Gamma}
\end{equation}
The macroscopic observable we are interested in calculating is the
overlap $m=\lim_{N\to\infty}\frac1N\sum_i\xi_i\hat{\xi}_i$ between
the plaintext and the Bayes Marginal Posterior Maximizer (MPM)
estimate of the plaintext $\hat{\xi}_i\equiv{\rm
sign}\sum_{\s_i=\pm}\s_i\ p(\s_i|\br)$ where $p(\s_i|\br)$ is the
microscopic state probability (\ref{eq:state_prob}). Disorder
averages $\langle \rangle_{\Gamma}$ are taken over the probability
distributions
(\ref{eq:prob_xi},\ref{eq:prob_zeta},\ref{eq:prob_wp},\ref{eq:prob_wc})
and over the distribution of the tensors $\mcA$ and $\mcG$ obeying
the constrains (\ref{eq:mcA1}-\ref{eq:mcG2}):
\begin{eqnarray}
\left\bra \mc{F}\left(\mcA\right)\right\ket_{\mcA,\{\tilde{C}_i,\tilde{L}_i\}}
\nonumber
&=&
\frac{1}{\mathcal{N}}
 \sum_{\{\mcA_{\LK\OL}\}}
 \prod_{i=1}^N\left\bra
 \delta\left[\sum_{\LK\OL/i\in\LK}\mcA_{\LK\OL}-\tilde{C}_{i}\right]\right\ket_{P(\tilde{C}_{i})}
\nonumber
\\
&&
\hspace{16mm}
\times\
 \prod_{j=1}^M
 \left\bra\delta\left[\sum_{\LK\OL/j\in\OL}\mcA_{\LK\OL}-\tilde{L}_{j}\right]
 \right\ket_{P(\tilde{L}_{j})} \mc{F}(\mcA)
\label{eq:mcA}
\\
\left\bra \mc{F}\left(\mcG\right)\right\ket_{\mcG}
&=&
\frac{1}{\mathcal{N}^\prime}
 \sum_{\{\mcG_{\LKP\OLP}\}}
 \prod_{i=1}^N
 \delta\left[\sum_{\LKP\OLP/i\in\LKP}\mcA_{\LKP\OLP}-C^\prime\right]\
\nonumber
\\
&&
\hspace{16mm}
\times\
 \prod_{j=1}^M
 \delta\left[\sum_{\LKP\OLP/j_1\in\OLP}\mcG_{\LKP\OLP}-1\right]
  \mc{F}(\mcG)
\label{eq:mcG}
\end{eqnarray}
where $\mc{N}$ and $\mc{N}^\prime$ are the corresponding normalisation constants.

The parameters $w_{\sigma},w_{\tau},F_{\sigma},F_{\tau}$ and
$\gamma$ describe the attack characteristics.

%--------------------------------------------------------------------%
\section{The Free Energy  and Decryption Observables}\label{sec:analysis}
%--------------------------------------------------------------------%

The calculation generally follows that of~\cite{kabashima,MKSV}.
To perform the various disorder averages we begin by invoking the
replica identity $\bra \log Z\ket=\lim_{n\to 0}\frac1n\log \bra
Z^n\ket$ and making the gauge transformations $\s_i\to\s_i\xi_i$,
$\tau_i\to\tau_i\zeta_i$,
$\mcA_{\LK\OL}\to\mcA_{\LK\OL}\prod_{i\in\LK}\xi_i\prod_{j\in\OL}\zeta_j$
and
$\mcG_{\LKP\OLP}\to\mcG_{\LKP\OLP}\prod_{i\in\LKP}\xi_i\prod_{j\in\OLP}\zeta_j$.
This will allow us to disentangle the variables $\{\xi,\zeta\}$
from expressions involving the tensors $\mcA$ and $\mcG$ in
(\ref{eq:Delta_A},\ref{eq:Delta_G}). Replacing the $\delta$
functions in (\ref{eq:mcA},\ref{eq:mcG}) by their integral
representations allows us to perform the tensor summations,
leading to:
\begin{eqnarray}
\lefteqn{\left\bra\, \Delta_{\mcA}(\bsigma,\btau),\Delta_{\mcG}(\bsigma,\btau)\,\right\ket=}
\nonumber
\\
&&
=\frac{1}{\mc{N}\mc{N}^\prime}
\oint\frac{\prod_{i=1}^NdZ_idX_i}{(2\pi)^{2N}}\
\oint\frac{\prod_{j=1}^MdY_jdV_j}{(2\pi)^{2M}}
\nonumber
\\
& & \times\ \prod_{i=1}^N \left\bra Z_i^{-(\tilde{C}_i+1)}
X_i^{-(C^\prime+1)}\right\ket_{P(\tilde{C}_i)}\ \prod_{j=1}^M
\left\bra Y_j^{-(\tilde{L}_j+1)} V_j^{-2}
 \right\ket_{P(\tilde{L}_j)}
\nonumber
\\
& &
\times\
e^{(\frac12)^n\sum_{m=0}^n\sum_{\bra \alpha_1\cdots\alpha_m\ket}
   \frac{1}{K!}\left(\sum_{i=1}^NZ_i\s_i^{\alpha_1}\cdots\s_i^{\alpha_m}\right)^K
   \frac{1}{L!}\left(\sum_{j=1}^MY_j\tau_j^{\alpha_1}\cdots\tau_j^{\alpha_m}\right)^L}
\nonumber
\\
& &
\times\
e^{(\frac12)^n\sum_{m=0}^n\sum_{\bra \alpha_1\cdots\alpha_m\ket}
   \frac{1}{K^\prime!}\left(\sum_{i=1}^N X_i\s_i^{\alpha_1}\cdots\s_i^{\alpha_m}\right)^{K^\prime}
   \left(\sum_{j=1}^M V_j\tau_j^{\alpha_1}\cdots\tau_j^{\alpha_m}\right)}
\label{eq:tensor_av1}
\end{eqnarray}
In the above expression we can now identify the following order parameters
\begin{equation}
q_{\alpha_1\cdots\alpha_m}=\sum_{i=1}^NZ_i\s_i^{\alpha_1}\cdots \s_i^{\alpha_m}
\hspace{10mm}
r_{\alpha_1\cdots\alpha_m}=\sum_{i=1}^NX_i\s_i^{\alpha_1}\cdots \s_i^{\alpha_m}
\end{equation}
\begin{equation}
t_{\alpha_1\cdots\alpha_m}=\sum_{j=1}^MY_j\tau_j^{\alpha_1}\cdots \tau_j^{\alpha_m}
\hspace{10mm}
u_{\alpha_1\cdots\alpha_m}=\sum_{j=1}^MV_j\tau_j^{\alpha_1}\cdots \tau_j^{\alpha_m}
\end{equation}
which we insert in (\ref{eq:tensor_av1}) via suitably defined
$\delta$ functions (giving rise to the Lagrange multipliers
$\hat{q}_{\alpha_1\cdots\alpha_m}$,
$\hat{r}_{\alpha_1\cdots\alpha_m}$ ,
$\hat{t}_{\alpha_1\cdots\alpha_m}$ and
$\hat{u}_{\alpha_1\cdots\alpha_m}$). To proceed with the
calculation one needs to assume a certain order parameter symmetry
for the above quantities and their conjugates for all $m>1$. The
simplest such assumption renders all replica $m$-tuples equivalent
and all order parameters within this replica symmetric scheme need
only depend on the number $m$. This effect can be described by the
introduction of suitably defined distributions, the moments of
which completely define the $m$-index order parameters
\begin{equation}
q_{\alpha_1\cdots\alpha_m}=q\int dx\ \pi(x)\ x^m
\hspace{10mm}
\hat{q}_{\alpha_1\cdots\alpha_m}=\hat{q}\int dx\ \hat{\pi}(x)\ x^m
\label{eq:pi}
\end{equation}
\begin{equation}
r_{\alpha_1\cdots\alpha_m}=r\int dy\ \rho(y)\ y^m
\hspace{10mm}
\hat{r}_{\alpha_1\cdots\alpha_m}=\hat{r}\int dy\ \hat{\rho}(y)\ y^m
\end{equation}
\begin{equation}
t_{\alpha_1\cdots\alpha_m}=t\int dx\ \phi(x)\ x^m
\hspace{10mm}
\hat{t}_{\alpha_1\cdots\alpha_m}=\hat{t}\int dx\ \hat{\phi}(x)\ x^m
\end{equation}
\begin{equation}
u_{\alpha_1\cdots\alpha_m}=u\int dy\ \psi(y)\ y^m
\hspace{10mm}
\hat{u}_{\alpha_1\cdots\alpha_m}=\hat{u}\int dy\ \hat{\psi}(y)\ y^m
\label{eq:psi}
\end{equation}
where all integrals are over the interval $[-1,1]$. The Nishimori
condition ($\beta=1$), which corresponds to MPM decoding~\cite{iba},
also ensures that this simplest replica-symmetric scheme is sufficient
to describe the thermodynamically dominant
state~\cite{nishimori,NS}. Furthermore, it is worthwhile mentioning
that extending the replica symmetric calculation to include the
one-step replica symmetry breaking ansatz is unlikely to modify the
location of the transition points identified under the
replica-symmetric ansatz, as has been recently shown in a similar
system~\cite{franz}. Using the above ansatz we perform the contour
integrals in (\ref{eq:tensor_av1}), and trace over the spin variables;
then, in the limit $n\to 0$ we obtain:
\begin{eqnarray}
\label{eq:free_energy}
-\beta f
&=&
{\rm Extr}\left\{
 -\CB J_{1a}[\pi,\hat{\pi}]-\frac{\CB L}{K}J_{1b}[\rho,\hat{\rho}]
-C^\prime J_{1c}[\phi,\hat{\phi}]-\frac{C^\prime }{K^\prime}J_{1d}[\psi,\hat{\psi}]
\right.
\\
&&
\left.
+\frac{\CB}{K}J_{2a}[\pi,\rho]+\frac{C^\prime}{K^\prime}J_{2b}[\phi,\psi]+J_{3a}[\hat{\pi},\hat{\phi}]
+\frac{\CB}{K}\frac{L}{\LB}J_{3b}[\hat{\rho},\hat{\psi}]\right\}
-\left(\frac{\CB}{K}+\frac{C^\prime}{K^\prime}\right)\log 2
\nonumber
\end{eqnarray}
where the extremization is taken over the distributions defined in
(\ref{eq:pi}-\ref{eq:psi}) and the various integrals
$J_{\star\star}$ are given by
\begin{equation}
J_{1a}[\pi,\hat{\pi}]=\int dxd\hat{x}\ \pi(x)\hat{\pi}(\hat{x})\ \log(1+x\hat{x})
\hspace{10mm}
J_{1b}[\rho,\hat{\rho}]=\int dyd\hat{y}\ \rho(y)\hat{\rho}(\hat{y})\ \log(1+y\hat{y})
\label{eq:J1a}
\end{equation}
\begin{equation}
J_{1c}[\phi,\hat{\phi}]=\int dxd\hat{x}\ \phi(x)\hat{\phi}(\hat{x})\ \log(1+x\hat{x})
\hspace{10mm}
J_{1d}[\psi,\hat{\psi}]=\int dyd\hat{y}\ \psi(y)\hat{\psi}(\hat{y})\ \log(1+y\hat{y})
\label{eq:J1c}
\end{equation}
\begin{equation}
J_{2a}[\pi,\rho]=\int [\prod_{k=1}^Kdx_k\ \pi(x_k) \prod_{\ell=1}^Ldy_\ell\ \rho(y_\ell)]\
   \log(1+\prod_k x_k\prod_\ell y_\ell)
\label{eq:J2a}
\end{equation}
\begin{equation}
J_{2b}[\phi,\psi] = \int dy\ \psi(y)\ [\prod_{k=1}^{K^\prime}dx_k\ \phi(x_k)]\
   \log(1+y\prod_k x_k)
\label{eq:J2b}
\end{equation}
\begin{eqnarray}
\lefteqn{J_{3a}[\hat{\pi},\hat{\phi}]
=
\int \prod_{c'=1}^{C^\prime}
  d\hat{\phi}(y_{c'})\left\{(1-\gamma)^C\left\bra\log \sum_{\lambda=\pm}[(1-c)+c\delta_{\lambda,1}]e^{\beta F_{\sigma}\xi
  \lambda}\prod_{c'}(1+y_{c'}\lambda)\right\ket_{c,\xi}\right.}
\nonumber
\\
& & \left. +\left\bra \int
[\prod_{c=1}^{\tilde{C}}d\hat{\pi}(x_c)] \left\bra\log
\sum_{\lambda=\pm}[(1-c)+c\delta_{\lambda,1}]e^{\beta F_{\sigma}\xi \s}
\prod_{c}(1+x_c\lambda)\prod_{c'}(1+y_{c'}\lambda)\right\ket_{c,\xi}\right\ket_{\tilde{C}}\right\}
\label{eq:J3a}
\end{eqnarray}
\begin{eqnarray}
\lefteqn{J_{3b}[\hat{\rho},\hat{\psi}] =\int dy~\hat{\psi}(y)\
\left\{(1-\gamma)^L
  \left\bra\log \sum_{\lambda=\pm}[(1-d)+d\delta_{\lambda,1}]
    e^{\beta F_{\tau}\zeta \lambda}(1+y\lambda)\right\ket_{d,\zeta}\right.}
\nonumber
\\
& &
\left.
+\left\bra\int[\prod_{\ell=1}^{\tilde{L}}d\hat{\rho}(x_\ell)]
 \left\bra\log \sum_{\lambda=\pm}[(1-d)+d\delta_{\lambda,1}]
    e^{\beta F_{\tau}\zeta \lambda}
\prod_{\ell}(1+x_\ell\lambda)(1+y\lambda)\right\ket_{d,\zeta}\right\ket_{\tilde{L}}\right\}
\label{eq:J3b}
\end{eqnarray}
where
\begin{equation}
\CB=\sum_{\tilde{C}=0}^C P(\tilde{C};C)\ \tilde{C} \hspace{10mm}
\LB=\sum_{\tilde{L}=0}^L P(\tilde{L};L)\ \tilde{L}
\end{equation}
Averages denoted $\bra \cdots \ket_{\tilde{C}}$ and $\bra \cdots
\ket_{\tilde{L}}$ are over the densities (\ref{eq:prob_C}) and
(\ref{eq:prob_L}) with $\tilde{C}=1,\ldots,C$ and $\tilde{L}=1,\ldots,L$.
Functional differentiation of (\ref{eq:free_energy}) with
respect to the densities of (\ref{eq:pi}-\ref{eq:psi}) results in
the following saddle point equations:
\begin{eqnarray}
\hat{\pi}(\hat{x})
&=&
\int[\prod_{k=1}^{K-1}dx_k\pi(x_k)\prod_{l=1}^{L}dy_l\rho(y_l)]\
   \delta\left[\hat{x}-\prod_{k=1}^{K-1}x_k\prod_{l=1}^Ly_l\right]
\label{eq:saddle_pihat}
\\
\hat{\rho}(\hat{y})
&=&
\int[\prod_{k=1}^{K}dx_k\pi(x_k)\prod_{l=1}^{L-1}dy_l\rho(y_l)]\
   \delta\left[\hat{y}-\prod_{k=1}^{K}x_k\prod_{l=1}^{L-1}y_l\right]
\\
\hat{\phi}(\hat{x})
&=&
\int dy\psi(y)\ [\prod_{k=1}^{K^\prime-1}dx_k\phi(x_k)]\
   \delta\left[\hat{x}-y\prod_{k=1}^{K^\prime-1}x_k\right]
\\
\hat{\psi}(\hat{y})
&=&
\int[\prod_{k=1}^{K^\prime}dx_k\phi(x_k)]\
   \delta\left[\hat{y}-\prod_{k=1}^{K^\prime}x_k\right]
\end{eqnarray}
and

\begin{eqnarray}
\lefteqn{\pi(x)=w_{\sigma}\ \delta[x-1]} \label{eq:saddle_pi}
\\
& & +\frac{(1-w_{\sigma})}{\CB}\left\bra\tilde{C} \int[
\prod_{c'=1}^{C^\prime}d\hat{\phi}(\hat{y}_{c'})
\prod_{c=1}^{\tilde{C}-1}d\hat{\pi}(\hat{x}_c)]
 \left\bra
 \delta\left(x-\tanh[\beta F_{\sigma}\xi+\sum_{c=1}^{\tilde{C}-1}\atanh(\hat{x}_c)
 +\sum_{c'=1}^{C^\prime}\atanh(\hat{y}_{c'})]\right)\right\ket_\xi
 \right\ket_{\tilde{C}}
\nonumber
\\
& &
\nonumber
\\
\lefteqn{\rho(x)=w_{\tau}\ \delta[x-1]} \label{eq:saddle_rho}
\\
& & +\frac{(1-w_{\tau})}{\LB} \left\bra\tilde{L}\int
d\hat{\psi}(\hat{y})\
   [\prod_{l=1}^{\tilde{L}-1}d\hat{\rho}(\hat{y}_l)]
 \left\bra \delta\left(x-\tanh[\beta F_{\tau}\zeta+\sum_{l=1}^{\tilde{L}-1}\atanh(\hat{x}_l)
                                 +\atanh(\hat{y})]\right)\right\ket_\zeta
 \right\ket_{\tilde{L}}
\nonumber
\\
& &
\nonumber
\\
\lefteqn{\phi(x)=w_{\sigma}\ \delta[x-1]} \label{eq:saddle_phi}
\\
& &
  +(1-w_{\sigma})\int\prod_{c'=1}^{C^\prime-1}d\hat{\phi}(y_{c'})
  \left\{(1-\gamma)^C\left\bra \delta\left(x-\tanh[\beta F_{\sigma}\xi+
   \sum_{c'=1}^{C^\prime-1}\atanh(\hat{y}_{c'})]\right)\right\ket_\xi\right.
\nonumber
\\
& &
\left.
+\left\bra
 \int[\prod_{c=1}^{\tilde{C}}d\hat{\pi}(\hat{x}_c)]
 \left\bra \delta\left(x-\tanh[\beta F_{\sigma}\xi+\sum_{c=1}^{\tilde{C}}\atanh(\hat{x}_c)
                                 +\sum_{c'=1}^{C^\prime-1}\atanh(\hat{y}_{c'})]\right)\right\ket_\xi
 \right\ket_{\tilde{C}}\right\}
\nonumber
\\
& &
\nonumber
\\
\lefteqn{\psi(x)=w_{\tau}\ \delta[x-1]} \label{eq:saddle_psi}
\\
& & +(1-w_{\tau})\left\{(1-\gamma)^L\left\bra\delta[x-\tanh(\beta
F_{\tau}\zeta)]\right\ket_{\zeta} +\left\bra
  \int[\prod_{l=1}^{\tilde{L}}d\hat{\rho}(\hat{x}_l)]
 \left\bra \delta\left(x-\tanh[\beta F_{\tau}\zeta+\sum_{l=1}^{\tilde{L}}\atanh(\hat{x}_l)]\right)\right\ket_\zeta
\right\ket_{\tilde{L}}\right\}
\nonumber
\end{eqnarray}
In general, the coupled set of equations
(\ref{eq:saddle_pihat})-(\ref{eq:saddle_psi}) are to be solved
numerically. Among the set of $\bsigma$ that satisfy equations
(\ref{eq:sparse-part}) and (\ref{eq:dense-part}) we choose the MPM
estimate of the plaintext $\hat{\xi}_i={\rm
sign}\sum_{\s_i=\pm}\s_i\ p(\s_i|\br)={\rm sign}\bra\s_i\ket$
(thermal average) by using Nishimori's condition (or
$\beta=1$)~\cite{nishimori}. Then, the overlap
$m=\lim_{N\to\infty}\frac1N\sum_i\xi_i\hat{\xi}_i$ becomes
\begin{eqnarray}
m &=& w_{\sigma}+(1-w_{\sigma})\int dh\ P(h)\ {\rm sign} (h)
\label{eq:overlap}
\\
P(h) & = & \int \prod_{c'=1}^{C^\prime}d\hat{\phi}(\hat{y}_{c'}) ]
\left\{ (1-\gamma)^C\left\bra\delta\left(h-\tanh[\beta
F_{\sigma}\xi+
   \sum_{c'=1}^{C'}\atanh(\hat{y}_{c'})]\right)\right\ket_{\xi}\right.
\nonumber
\\
& & +\left.
\left\bra\int[\prod_{c=1}^{\tilde{C}}d\hat{\pi}(\hat{x}_c)]
\left\bra\delta\left(h-\tanh[\beta
F_{\sigma}\xi+\sum_{c=1}^{\tilde{C}}\atanh(\hat{x}_c)
+\sum_{c'=1}^{C^\prime}\atanh(\hat{y}_{y'})]\right)\right\ket_{\xi}\right\ket_{\tilde{C}}\right\}
\end{eqnarray}
from which it can be seen that the perfect (ferromagnetic)
solution $m=1$ is achieved when $w_{\sigma}=1$ (complete knowledge
of the solution) or when $\hat{\phi}(x)=\delta[x-1]$. This also
implies that all densities involved in (\ref{eq:free_energy})
$\lambda(x)=\{\pi(x),\ldots,\hat{\psi}(x)\}$ acquire the form
$\lambda(x)=\delta[x-1]$ giving a free energy of the form
\begin{equation}
f_{FM}=\left(\frac{C^\prime}{K^\prime}-\frac{C}{K}\right)\log
2-\frac{C}{K}\beta F_{\tau}\bra \zeta\ket_{\zeta}
\end{equation}
The physical meaning of the terms $w_\star\,\delta[x-1]$ in
(\ref{eq:saddle_pi}-\ref{eq:saddle_psi}) is that the acquired
microscopic knowledge gives  a probabilistic weight at the
ferromagnetic state. The state $m=0$ is obtained if
$w_{\sigma}=F_{\sigma}=0$ and
$\hat{\pi}(x)=\hat{\phi}(x)=\delta[x]$ (paramagnetic solution).

%--------------------------------------------------------------------%
\section{Phase Diagrams} \label{sec:results}
%--------------------------------------------------------------------%

In this section we obtain numerical solutions for various attack
scenarios. In all cases studied we assume an unbiased plaintext
($p_{\sigma}=1/2,~F_{\sigma}=0$); for brevity we refer to the
remaining bias parameter, the corruption level denoted $p_{\tau}$
in previous sections, simply as $p$. All experiments have been
carried out using a regular cryptosystem with $K=L=2$, being the
original cryptosystem suggested in~\cite{KMS}. In principle, one
can use any set of regular or irregular matrices, provided one
identifies the corresponding dynamical transition point. However,
having been thoroughly studied previously, the current
construction serves as a particularly suited benchmark.

Solving the coupled equations
(\ref{eq:saddle_pihat}-\ref{eq:saddle_psi}) we typically observe that
for sufficiently small values of $p$ the ferromagnetic state $m=1$ is
the only stable solution whereas at a corruption value that marks the
dynamical (spinodal) transition $p_s$, an exponential number of
solutions with $m\neq 1$ are created (either suboptimal ferromagnetic
or paramagnetic, depending on the values of $(K,C,L)$). For all
$p>p_s$ perfect decryption will be difficult to obtain. This
transition also defines the corruption level below which an
unauthorized attacker, that have acquired partial information of the
secret keys, will be successful.

We will concentrate on two main attacks: (i) The attacker has
partial knowledge of the keys (primarily the matrix $B$). (ii) The
attacker has partial microscopic knowledge of the plaintext and/or
corruption vector.

\begin{center}
\begin{figure}[t]
\setlength{\unitlength}{1.1mm}
\begin{picture}(120,55)
\put( 30, 5){\epsfysize=50\unitlength\epsfbox{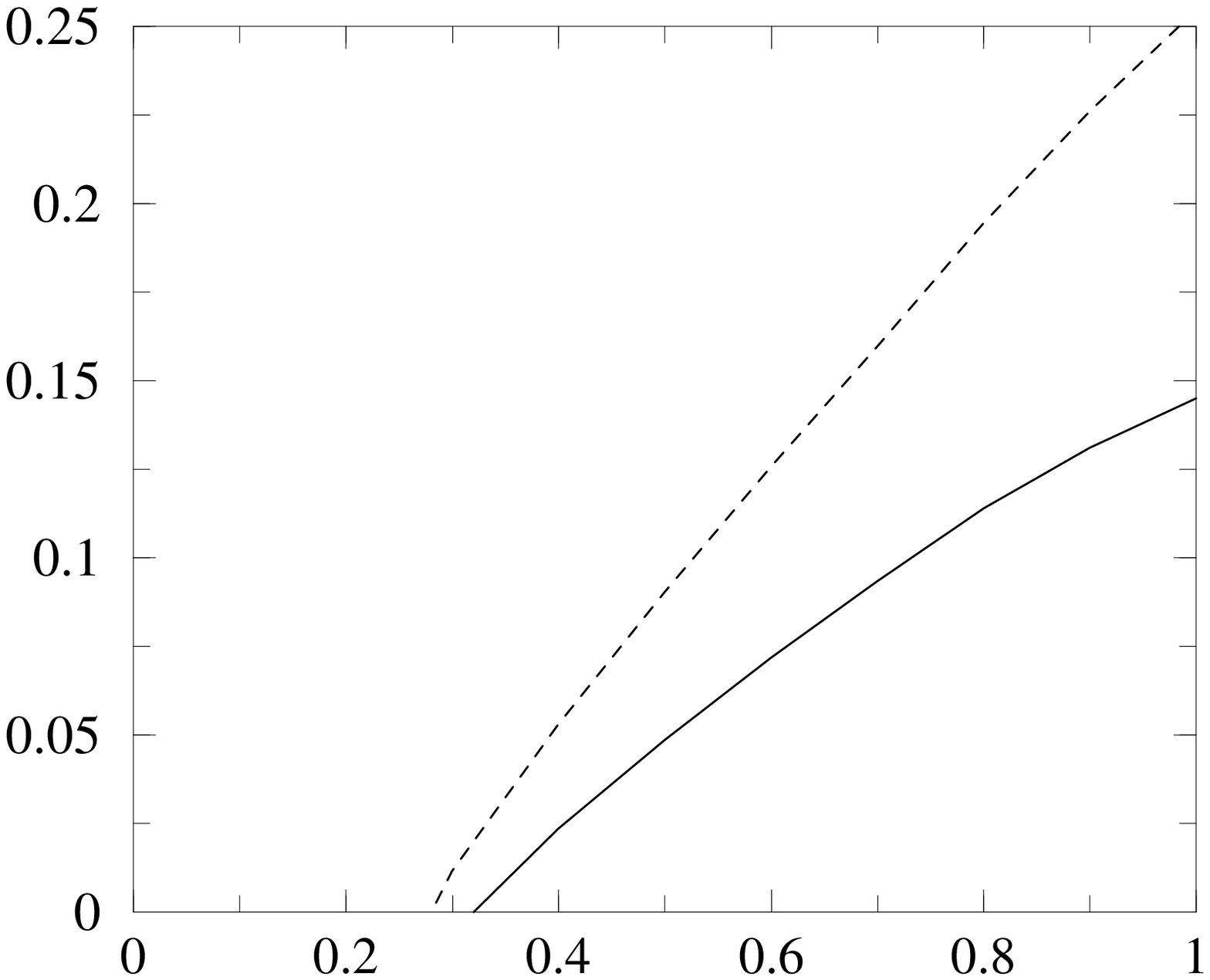}}
\put(20,35){\large $p_s$}
\put(65,0) {\large $\gamma$}
\put(75,15){\large $m=1$}
\put(43,37){\large $|m|<1$}
\end{picture}
\caption{Phase diagram of the spinodal corruption-rate against the
fractional knowledge of the private key $\gamma$ for a
$(K,C,L)=(2,6,2)$ cryptosystem for $(w_{\sigma},w_{\tau})=(0,0)$
(solid line) and $(0.2,0.2)$ (dashed line). Microscopic knowledge
of the plaintext and the corrupting vector enlarges the perfect
decryption area, as expected.} \label{fig:PhaseD1}
\end{figure}
\end{center}

In figure \ref{fig:PhaseD1} we present a phase diagram describing
regions with perfect ($m=1$) or partial/null ($|m|<1$) decryption
success as evaluated from solving equations (\ref{eq:free_energy})
and (\ref{eq:overlap}). We plot the dynamical transition
corruption level $p_s$ as a function of the private key fractional
knowledge $\gamma$ for different values of $w_{\sigma}$ and
$w_{\tau}$ (we have set $p_{\sigma}=1/2$ which corresponds to an
`unbiased' plaintext). In the limit $\gamma=0$ (i.e., no knowledge
of the matrices), while $m=1$ may be a stable solution, the
decryption dynamics is fully dominated by $|m|<1$ states. For
$\gamma=1$ the cryptosystem describes a specific MN code and
perfect decryption can occur below $p_s$.

\begin{center}
\begin{figure}[bh]
\setlength{\unitlength}{1.1mm}
\begin{picture}(120,55)
\put( -5, 5){\epsfysize=50\unitlength\epsfbox{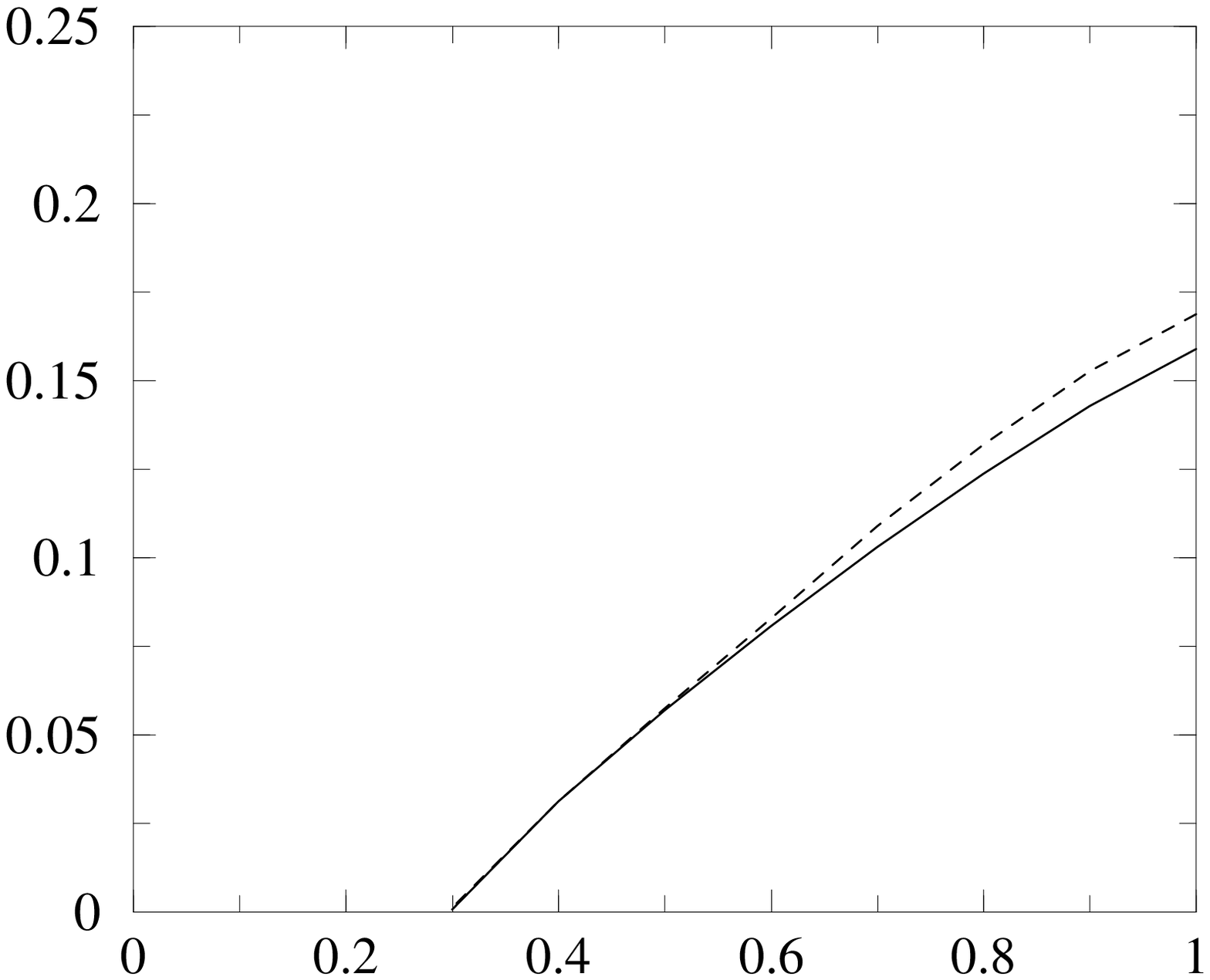}}
\put( 70, 5){\epsfysize=50\unitlength\epsfbox{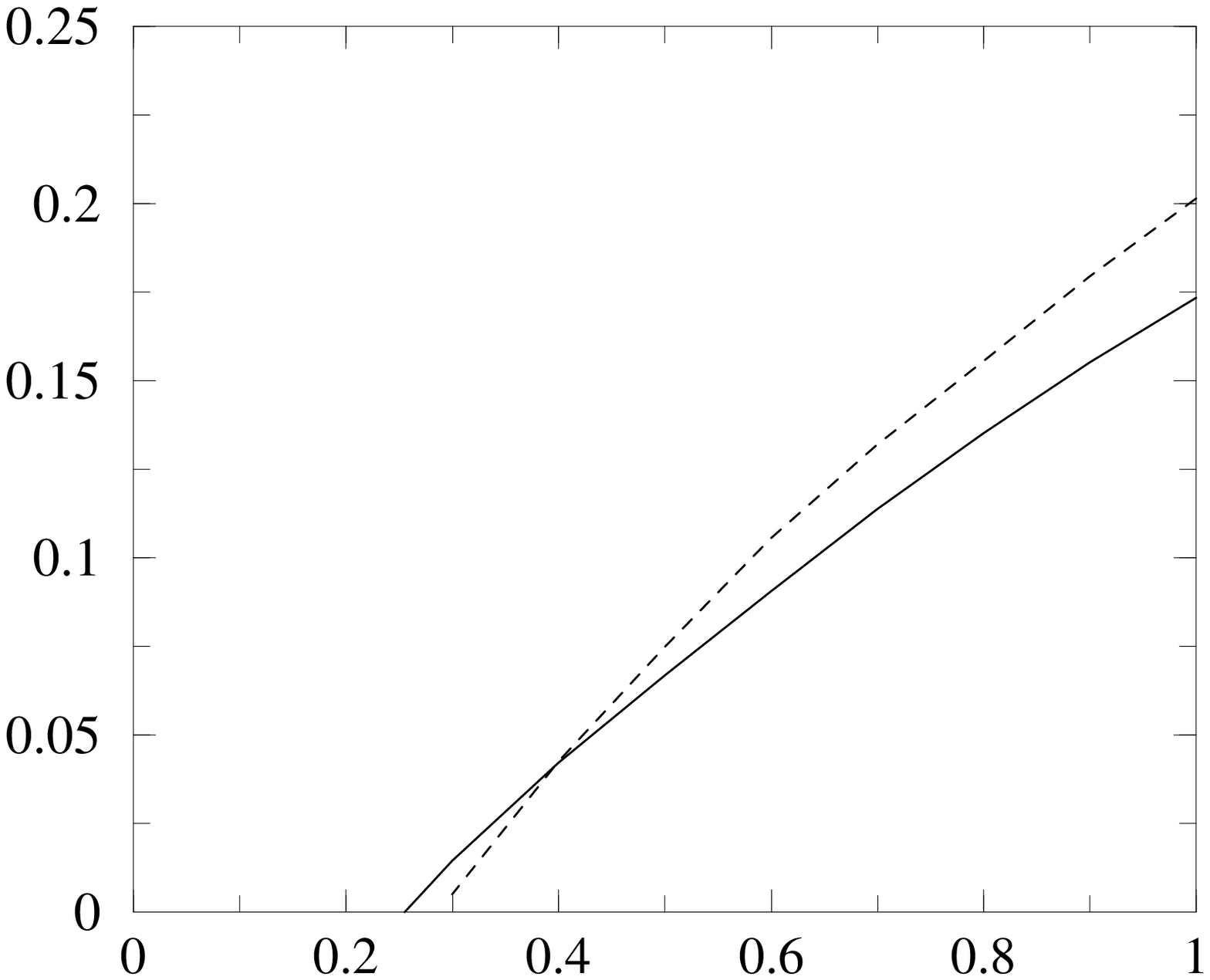}}
\put(-10,30){\large $p_s$}
\put(35,0) {\large $\gamma$}
\put(65,30){\large $p_s$}
\put(105,0) {\large $\gamma$}
\put(43,15){\large $m=1$}
\put(112,15){\large $m=1$}
\put(15,35){\large $|m|<1$}
\put(85,35){\large $|m|<1$}
\end{picture}
\caption{Phase diagrams of the spinodal corruption-rates against
the fractional knowledge of the private key $\gamma$ for a
$(K,C,L)=(2,6,2)$ cryptosystem. Left picture:
$(w_{\sigma},w_{\tau})=(0.1,0)$ (solid line) and $(0,0.1)$ (dashed
line). Right picture: $(w_{\sigma},w_{\tau})=(0.2,0)$ (solid line)
and $(0,0.2)$ (dashed line). For sufficiently large
$\gamma$-values microscopic knowledge of the corrupting vector
becomes more important to the unauthorized user than that of the
plaintext; this effect becomes more emphasized as the fraction of
known bits increases.} \label{fig:PhaseD2}
\end{figure}
\end{center}

\begin{center}
\begin{figure}[t]
\setlength{\unitlength}{1.1mm}
\begin{picture}(120,55)
\put( -5, 5){\epsfysize=50\unitlength\epsfbox{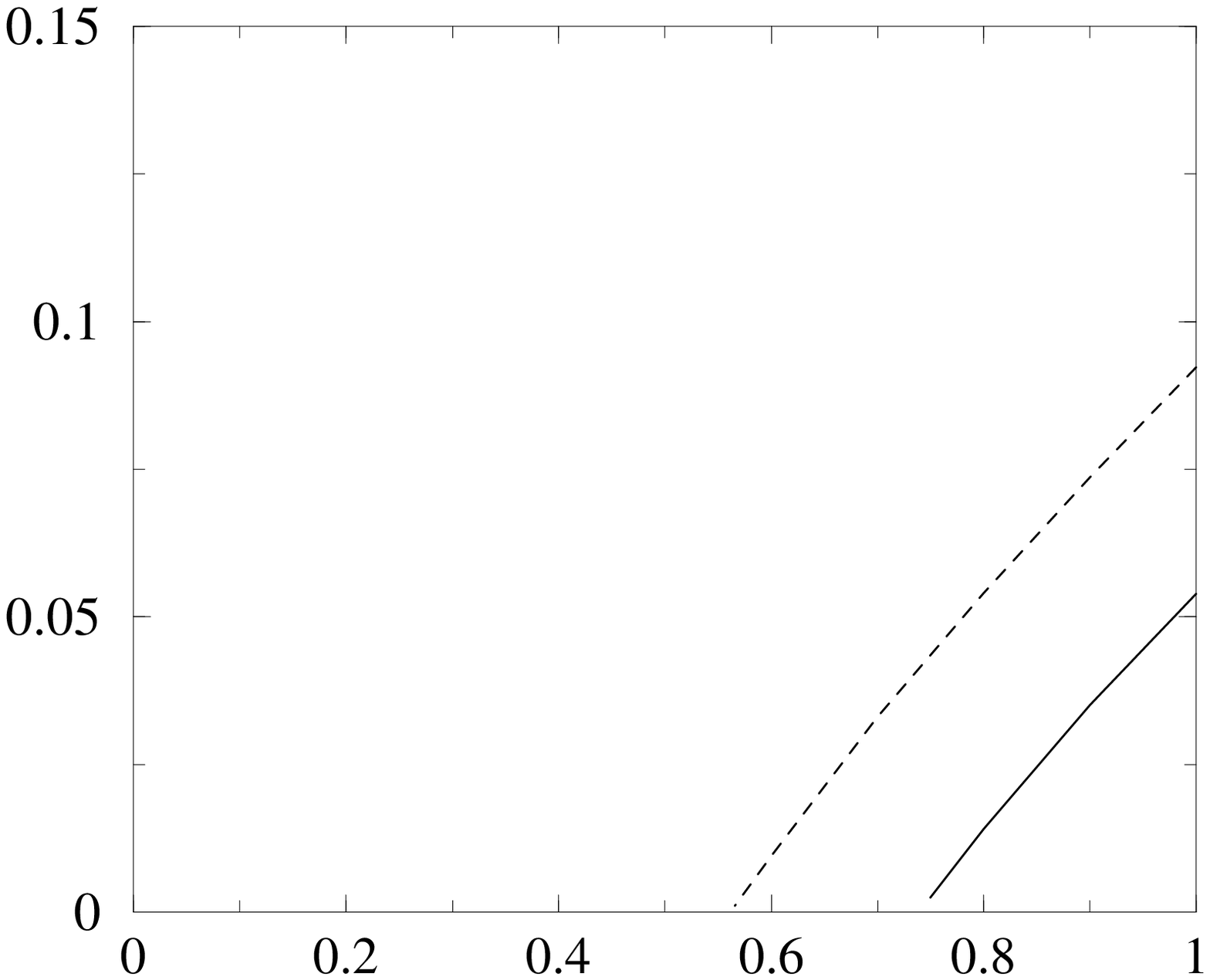}}
\put( 70, 5){\epsfysize=50\unitlength\epsfbox{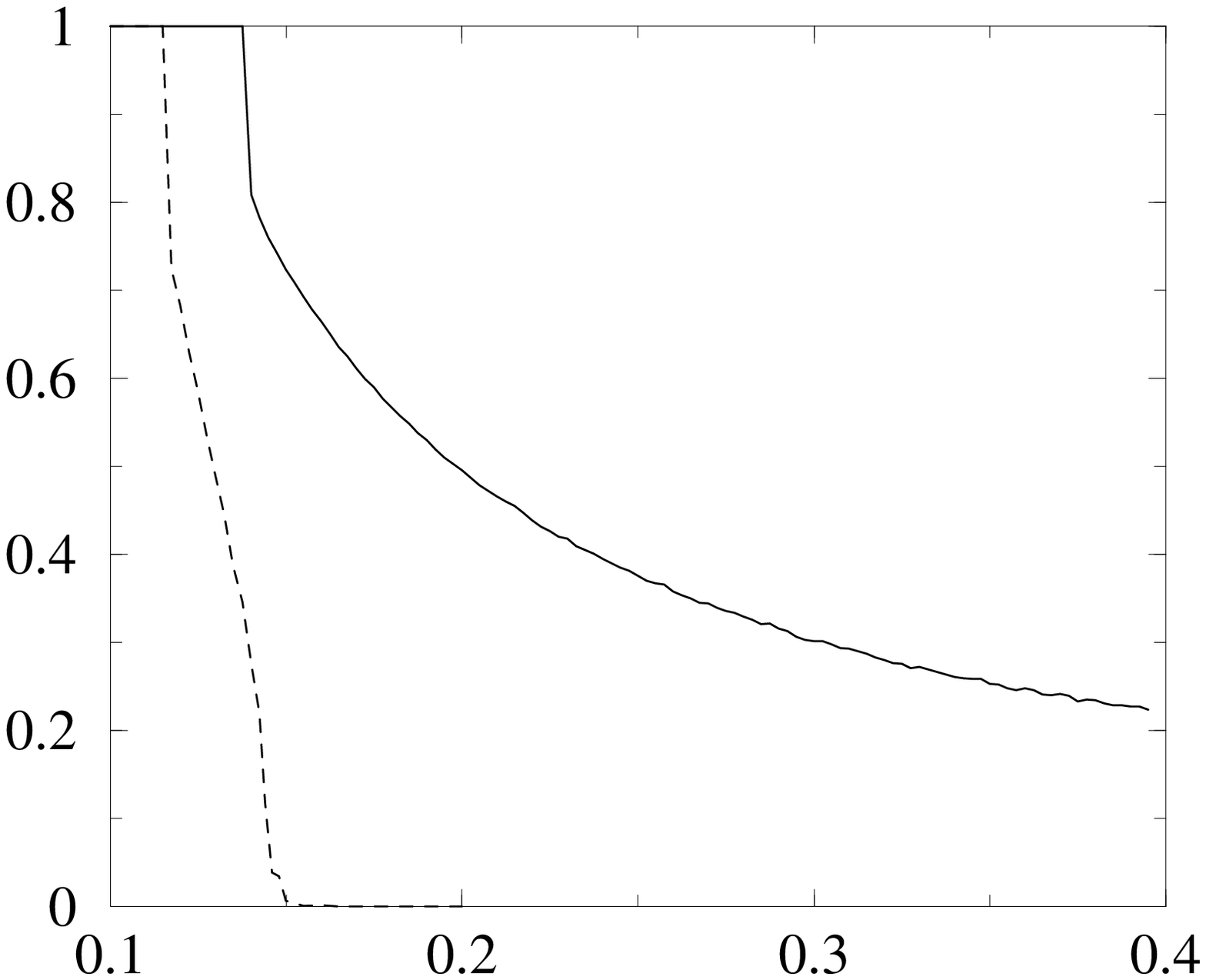}}
\put(-7,30){\large $p_s$}
\put(35,0) {\large $\gamma$}
\put(63,30){\large $m$}
\put(105,0) {\large $p$}
\put(47,13){\small $m=1$}
\put(15,35){\large $|m|<1$}
\end{picture}
\caption{Left: Comparison between two different
cryptosystems with $(K,C,L)=(2,3,2)$ (solid line) and
$(K,C,L)=(2,4,2)$ (dashed line). Smaller $C$-values correspond to
higher rate codes and lead to smaller regions in parameter space
where perfect decryption is possible. Right: Overlap $m$ as
function of the corrupting-rate $p$ obtained from equation
(\ref{eq:overlap}) for a $(K,C,L)=(2,6,2)$ cryptosystem and along
the line $\gamma=0.8$ for $(w_{\sigma},w_{\tau})=(0.2,0)$ (solid
line) and $(w_{\sigma},w_{\tau})=(0,0)$ (dashed line).}
\label{fig:overlap}
\end{figure}
\end{center}

The interaction between the sparsely ({\ref{eq:sparse-part}) and
densely ({\ref{eq:dense-part}) connected decryption components is
non-linear and non-trivial; however, as a first approximation one
can view the fractional matrix knowledge $\gamma$ as changing the
effective sparse component, which is the main contributor in the
decryption process. To that end $\gamma$ will have a direct impact
on the effective code rate $N/(M\gamma)$, the average connectivity
$\gamma C$ and the connectivity distribution. It is clear that at
an effective code rate 1 ($\gamma=N/M=1/3$ in the case of the
parameters used in figure \ref{fig:PhaseD1}) decryption is even
not theoretically feasible.  The reason figure~\ref{fig:PhaseD1}
points to a possibility of decryption below this value is due to
additional information brought in by the dense components we
ignored in this simplistic description.

We also examined the effect of prior microscopic knowledge of the
plaintext/corrupting vector ($w_{\sigma},w_{\tau}>0$) on the area
of perfect decryption; which clearly increases with the knowledge
provided, as expected. Also this can be viewed as a change to the
effective code rate. This time, the partial microscopic knowledge
of either plaintext or corrupting vector (or both) serves to
reduce the effective number of variables and hence the code rate
itself; lower code rate will typically allow for perfect
decryption in worse corruption conditions as can be seen in
figure~\ref{fig:PhaseD1}

To understand the implication of these results let us assume using
the cryptosystem described in figure~\ref{fig:PhaseD1} at a
corruption level chosen of $p=0.1$ (which is chosen much smaller
that $p_s$ to increase the decryption reliability). In this case
knowing about $70\%$ of the matrices (secret keys) will be
sufficient for decrypting the ciphertext. True, there is still a
need to know the dense matrix $D^{-1}$ for extracting the
plaintext itself and the exposed fraction of the secret key is
significant; but still there is a weakness that may be exploited
by a skillful attacker.

To compare the importance of prior microscopic knowledge of
plaintext versus that of the corrupting vector we plotted in
figure \ref{fig:PhaseD2} the phase diagram for
$(w_{\sigma},w_{\tau})=\{(0.1,0),(0.2,0)\}$ and
$(w_\s,w_\tau)=\{(0,0.1),(0,0.2)\}$ which describe two
complementary scenarios (left and right figures respectively). The
effect is quite similar, taking into account the information
provided by the two vectors (the plaintext is unbiased but of
length $N$ while the corruption vector is biased but of length
$M$). For high $\gamma$-values microscopic knowledge of the
corrupting vector becomes more informative than that of the
plaintext, an effect which becomes more emphasized as the fraction
of known bits increases.

In figure~\ref{fig:overlap} we compare two cryptosystems with
$(K,C,L)=(2,4,2)$ and $(K,C,L)=(2,3,2)$ for
$(w_{\sigma},w_{\tau})=(0,0)$. We see that smaller $C$ values (i.e.,
higher code rates) will reduce the area of perfect decryption. On the
one hand, this will increase the secret information required for
perfect decryption at each corruption level; on the other hand it will
reduce the corruption level that can be used and will expose the
cryptosystem to attacks based on an exhaustive search of corruption
vectors.

The security of a cryptosystem may be compromised without a full
recovery of the plaintext; also partial recovery of the plaintext
may pose a significant threat. To study the effect of partial
knowledge of the matrices and plaintext on the ability to obtain
high overlap between the decrypted ciphertext and plaintext, we
conducted several experiments, an example of which appears in
figure \ref{fig:overlap}. Here we show the overlap obtained $m$ as
function of the corruption-rate $p$ for a specific cryptosystem
$(K,C,L)=(2,6,2)$ along the line $\gamma=0.8$ and for two
different choices of $w_{\sigma}$. Prior to the dynamical
transition points  both ciphertexts are decrypted perfectly; this
corresponds to corruption and partial knowledge levels below the
solid and dashed lines of figure~\ref{fig:PhaseD1}.

Above the dynamical transition point, new suboptimal solutions are
created and the overlap value obtained deteriorates with the
corruption level. However, the two different choices of
$w_{\sigma}$-values lead to two different deterioration patterns:
while overlap in the system with no microscopic knowledge of the
plaintext deteriorates very rapidly, the system with
$w_{\sigma}=0.2$ provides solutions with high overlap values even
if the corruption is high. As a consequence, we see that the
effect of microscopic knowledge goes beyond a shift in the
dynamical transition point; it also influences decryption beyond
that point (in fact, it goes even beyond Shannon's limit).

%--------------------------------------------------------------------%
\section{Basin of attraction}\label{sec:BOA}
%--------------------------------------------------------------------%
%
The increasingly narrowing basin of attraction for the ferromagnetic
solution, as the connectivity values $K,C$ and $L\to\infty$, is
central to the security level offered by the cryptosystem. The effect
has been reported in a number of papers in the statistical
physics~\cite{KMS,kanter} and information-theory~\cite{mackay}
literature; in this section we will show that the basin of attraction
shrinks as the connectivity increases, to a value of $O(1/K)$ as
$K,C\to\infty$.

To provide a rough evaluation of the basin of attraction
(BOA) for obtaining the ferromagnetic solution we focus on
Eq.~(\ref{eq:MN}) in the limit $K,C \to \infty$. BOA clearly
depends on the algorithm used;  here we focus on the Belief
Propagation (BP) algorithm, which is empirically known to be the
best practical algorithm for solving problems of the current type.
As far as we explored, no other schemes such as the naive mean
field and the Belief Revision algorithms exhibit better
performance than BP, which implies that our consideration on BP is
at least of a certain practical significance (Survey
Propagation~\cite{MPZ} has not yet been tested for these systems).

Let us represent prior knowledge on plain text $\bxi$ and noise
$\bzeta$ (in {Ising spin representation}) as the {\em prior
probabilities}
\begin{eqnarray}
P_i^{o}(\sigma_i)&=&\frac{\exp(F_{\sigma i} \sigma_i)}{2 \cosh
(F_{\sigma i})},
\label{eq:prior_sigma} \\
P_j^{o}(\tau_j)&=&\frac{\exp(F_{\tau j}\tau_j)}{2 \cosh (F_{\tau
j})}, \label{eq:prior_tau}
\end{eqnarray}
respectively. Here, the parameters $F_{\sigma i}$ and $F_{\tau j}$
express confidence of the prior knowledge per variable, which is a
generalization of the global prior terms $F_{\sigma}, F_{\tau}$
used earlier. Notice that this representation includes the case
that certain bits are completely determined by setting $|F_{\sigma
i}| \mbox{(or $|F_{\tau j}|$)} \to \infty$, enabling us to cover
various scenarios. In the following, we assume that the fraction
of completely determined bits is less than $1$ when $N,M \to
\infty$. Given prior probabilities (\ref{eq:prior_sigma}) and
(\ref{eq:prior_tau}), and the indicator function $\Delta
(\bsigma,\btau;\bxi,\bzeta,{\cal A})$ which is the alternative to
parity check equation (\ref{eq:MN}), the Bayesian framework
provides the {\em posterior  probability }
\begin{eqnarray}
P^{post}(\bsigma,\btau)=\frac{\Delta
(\bsigma,\btau;\bxi,\bzeta,{\cal A}) \prod_{i=1}^N P^{o}_i
(\sigma_i) \prod_{j=1}^M P^{o}_j (\tau_i)}{Z},
\label{eq:posterior}
\end{eqnarray}
where $Z$ is the normalization constant. Using
Eq.~(\ref{eq:posterior}), one can determine the best possible action
for minimizing the expected value of a given cost
function~\cite{iba}. As a cost function, we select here the Hamming
distance between the correct plain text $\bxi$ and its estimates
$\hat{\bxi}$, $L(\hat{\bxi},\bxi)=N-\sum_{i=1}^N \hat{\xi}_i \xi_i$;
this selection naturally offers the maximizer of posterior marginal
(MPM) decoding $\hat{\xi_i}={\rm sign}(m^\sigma_i)$ as the optimal
estimation strategy, where
\begin{equation}
m^{\sigma}_{i}=\sum_{\bsigma,\btau} \sigma_i~ P^{post}(\bsigma,\btau),
\label{eq:post_mean}
\end{equation}
is the average of spin $\sigma_i$ over the posterior probability
and ${\rm sign}(x)=1$ for $x >0$ and $-1$, otherwise.

Computational cost for an exact evaluation of the spin average
(\ref{eq:post_mean}) increases as $O(2^{N+M})$, which implies that
MPM decoding is practically difficult. An alternative approach is
to resort to an approximation such as BP. In the current case,
this means to iteratively solving the coupled equations (for
details of the derivation see~\cite{mackay,MKSV})
\begin{eqnarray}
\hat{m}^{\sigma}_{\mu i}&=&J_\mu \prod_{l \in {\cal
L}^{\sigma}(\mu) \backslash i} m_{\mu l}^{\sigma} \prod_{j \in
{\cal L}^{\tau}(\mu)} m_{\mu j}^{\tau}, \quad \hat{m}^{\tau}_{\mu
j}=J_\mu \prod_{l \in {\cal L}^{\sigma}(\mu)} m_{\mu l}^{\sigma}
\prod_{k \in {\cal L}^{\tau}(\mu) \backslash j } m_{\mu k}^{\tau},
\label{eq:horizontal} \\
m^{\sigma}_{\mu i}&=&\tanh(F_{\sigma i} + \sum_{\nu \in {\cal
M}^{\sigma}(i) \backslash \mu} \atanh (\hat{m}^{\sigma}_{\nu i})
), \quad m^{\tau}_{\mu j}=\tanh(F_{\tau j} + \sum_{\nu \in {\cal
M}^{\tau}(j) \backslash \mu} \atanh (\hat{m}^{\tau}_{\nu j}) ),
\label{eq:vertical}
\end{eqnarray}
where $J_\mu \equiv \left (\prod_{l \in {\cal L}^{\sigma}(\mu)}
\xi_l \prod_{j \in {\cal L}^{\tau}(\mu)} \zeta_j \right )$, ${\cal
L}^{\sigma}(\mu)$ and ${\cal L}^{\tau}(\mu)$ are the sets of indices
of non-zero elements in $\mu$th row of $A$ and $B$, respectively,
and ${\cal M}^{\sigma}(i)$ and ${\cal M}^{\tau}(j)$ are similarly
defined for columns of $A$ and $B$, respectively.  ${\cal
L}^{\sigma}(\mu) \backslash i$ denotes a set of indices in ${\cal
L}^{\sigma}$ other than $i$, and similarly for other symbols.
%DS - change 7/5/03
The variables $m^{\sigma/\tau}_{\mu i}$ and
$\hat{m}^{\sigma/\tau}_{\mu i}$ represent pseudo posterior
averages of $\sigma_i$ (or $\tau_{j}$) when the $\mu$th check
$J_\mu$ is left out, and the influence of a newly added $J_\mu$ on
$\sigma_i$ (or $\tau_{j}$), respectively (see~\cite{mackay,MKSV}
for details). Using $\hat{m}^{\sigma}_{\mu i}$, the posterior
average $m_i^{\sigma}$ is obtained as
\begin{eqnarray}
m_i^{\sigma}=\tanh(F_{\sigma i} + \sum_{\mu \in {\cal
M}^{\sigma}(i) } \atanh (\hat{m}^{\sigma}_{\mu i}) ).
\label{eq:full}
\end{eqnarray}

Let us investigate the condition necessary for finding the correct
solution by iterating Eqs.(\ref{eq:horizontal}) and
(\ref{eq:vertical}) in the limit $K,C \to \infty$. For this
purpose, we first employ the gauge transformation $\xi_{i}
m^{\sigma}_{\mu i} \to m^{\sigma}_{\mu i}$, $\xi_{i}
\hat{m}^{\sigma}_{\mu i} \to \hat{m}^{\sigma}_{\mu i}$, $\zeta_{j}
m^{\tau}_{\mu j} \to m^{\tau}_{\mu j}$, $\zeta_{j}
\hat{m}^{\tau}_{\mu j} \to \hat{m}^{\tau}_{\mu j}$ and $J_\mu
\left (\prod_{l \in {\cal L}^{\sigma}(\mu)} \xi_l \prod_{j \in
{\cal L}^{\tau}(\mu)} \zeta_j \right ) \to 1$. This decouples the
quenched random variables $\xi_i$ and $\zeta_j$ from
Eq.(\ref{eq:horizontal}), as $J_\mu$ becomes independent of the
quenched variables, and the BP equations can be expressed as
\begin{eqnarray}
\hat{m}^{\sigma}_{\mu i}&=& \prod_{l \in {\cal L}^{\sigma}(\mu)
\backslash i} m_{\mu l}^{\sigma} \prod_{j \in {\cal
L}^{\tau}(\mu)} m_{\mu j}^{\tau}, \quad \hat{m}^{\tau}_{\mu j}=
\prod_{l \in {\cal L}^{\sigma}(\mu)} m_{\mu l}^{\sigma} \prod_{k
\in {\cal L}^{\tau}(\mu) \backslash j } m_{\mu k}^{\tau},
\label{eq:horizontal2} \\
m^{\sigma}_{\mu i}&=&\tanh(F_i^{\sigma} \xi_i + \sum_{\nu \in
{\cal M}^{\sigma}(i) \backslash \mu} \atanh (\hat{m}^{\sigma}_{\nu
i}) ), \quad m^{\tau}_{\mu j}=\tanh(F_j^{\tau} \zeta_j + \sum_{\nu
\in {\cal M}^{\tau}(j) \backslash \mu} \atanh (\hat{m}^{\tau}_{\nu
j}) ). \label{eq:vertical2}
\end{eqnarray}
The expression of the correct solution is also converted to $m_{\mu
i}^{\sigma}=1$ and $m_{\mu j}^{\tau}=1$. Notice that any state which
is characterized by decreasing absolute values $|m_{\mu i}^{\sigma}| <
1 - \varepsilon$ and $|m_{\mu i}^{\tau}| < 1 - \varepsilon$ for an
arbitrary fixed positive number $\varepsilon > 0$ is attracted to a
locally stable solution $\hat{m}^{\sigma}_{\mu i} \sim 0$,
$\hat{m}^{\tau}_{\mu j} \sim 0$, $m^{\sigma}_{\mu
i}=\tanh(F_i^{\sigma} \xi_i)$ and $m^{\tau}_{\mu j}=\tanh(F_j^{\tau}
\zeta_j)$ for $K \to \infty$ in a single update since products on the
right hand sides of Eq.~(\ref{eq:horizontal2}) vanish. To provide a
rough evaluation of the BOA for the correct (ferromagnetic) solution
$m_{\mu i}^{\sigma}=1$ and $m_{\mu j}^{\tau}=1$, let us assume that
$m^{\sigma}_{\mu i}$ and $m^{\tau}_{\mu j}$ are randomly distributed
at $1-\varepsilon(K)$ and $-(1-\varepsilon(K))$ with probabilities
$1-p(K)$ and $p(K)$, respectively, where $\varepsilon(K)$ and $p(K)$
are small parameters to characterize the BOA for a large $K$. Under
this assumption, $\hat{m}^{\sigma}_{\mu i}$ and $\hat{m}^{\sigma}_{\mu
j}$ are distributed at $ \pm (1-\varepsilon(K))^{K+L} \sim \pm
(1-\varepsilon(K))^{K} $ with probability $(1 \pm (1- 2p(K))^{K+L})/2
\sim (1 \pm (1-2 p(K))^{K})/2$, respectively. If either
$(1-\varepsilon(K))^K$ or $(1-2p(K))^K$ is negligible, the absolute
values of $m_{\mu i}^{\sigma}$ and $m_{\mu j}^{\tau}$ become
sufficiently smaller than $1$, and therefore, the state is trapped in
a locally stable solution in the second iteration~\cite{footnote}.
This implies that the critical condition is given by $\varepsilon(K)
\sim O(1/K)$ and $p(K) \sim O(1/K)$ for large $K$. In terms of the
macroscopic overlap, this means $m_{cr}^0\approx 1-O(1/K)$.

%--------------------------------------------------------------------%
\section{Reliability}\label{sec:reliability}
%--------------------------------------------------------------------%

Unlike most of the commonly used cryptosystems which are based on
a deterministic decryption procedure, the current cryptosystem
relies on a probabilistic decryption process. The evaluation of
decryption success for an \emph{authorized} user is therefore as
important as assessing the level of robustness against attacks.

In practical scenarios, decryption success generally depends on
the plaintext size. Analysis of finite size effects in the belief
propagation based decryption procedure is  difficult. A principled
alternative that we pursue here is based on evaluating the {\em
average error exponent} of the current cryptosystem; this provides
the expected error-level at any given corruption level when
maximum likelihood decoding is employed, and therefore represents
a lower bound to the expected error-rate. Moreover, the corruption
levels employed are far below the critical (thermodynamic)
transition point, we therefore {\em assume} that belief
propagation decryption will provide similar performance to maximum
likelihood decoding; clearly, the lower bound will become looser
as we get close to the dynamical transition point.

The average block error rate $P_B(p)$ (i.e., erroneous decrypted plaintexts)
takes the form
\begin{equation}
 P_B(p) = e^{-M E(p)} \ ,
\end{equation}
where $E(p)$ is the average error exponent per noise level $p$ and
$M$ the length of the ciphertext (in the particular case of LDPC
codes we assume that short loops, which contribute polynomially to
the block error probability~\cite{miller}, have been removed). The
quantity $P_B(p)$ represents the probability by which candidate
solutions $\{\bsigma,\btau\}$ are drawn from the set of those
satisfying equation (\ref{eq:sparse-part}) (with $\gamma=1$;
authorized decryption) other than the ones corresponding to the
true plaintext and corrupting vector, $\bsigma=\bxi$ and
$\btau=\bzeta$, respectively.  To evaluate this probability we
introduce the indicator function
\begin{equation}
\Psi(\Gamma)=
\lim_{\beta\to\infty}\lim_{\lambda_{1,2}\to\pm\lambda}
\left[Z_1^{\lambda_1}(\Gamma;\beta_1)\
Z_2^{\lambda_2}(\Gamma;\beta_2)\right]_{\beta_1=\beta_2=\beta}
\label{eq:indic_function}
\end{equation}
where $\Gamma=\{\bxi,\bzeta,\mcA\}$ collectively denotes the set of
quenched variables. The power $\lambda\in[0,1]$ is used in conjunction
with the partition functions
\begin{equation}
Z_1(\Gamma;\beta_1)=
\sum_{\bsigma\neq\bxi}\sum_{\btau\neq \bzeta}
e^{-\beta_1 H(\bsigma,\btau)}
\hspace{10mm}
Z_2(\Gamma;\beta_2)=
\sum_{\bsigma}\sum_{\btau}
e^{-\beta_2 H(\bsigma,\btau)}
\label{eq:Z1andZ2}
\end{equation}
to provide an indicator function as explained below. The
Hamiltonian $H(\bsigma,\btau)$ is given by (\ref{eq:hamiltonian})
and the trace over spin variables is restricted to those
configurations satisfying equation (\ref{eq:sparse-part}). The
above partition functions $Z_1$ and $Z_2$ differ only in the
exclusion of the true plaintext and corrupting vector in the trace
over variables; this enables us to identify instances where the
maximum likelihood decoder chooses solutions that do not match the
true (quenched variable) vectors. The Hamiltonian
(\ref{eq:hamiltonian}) is proportional to the magnetizations
$m_\sigma(\bsigma)=\frac1N\sum_i\s_i$ and
$m_\tau(\btau)=\frac1M\sum_i\tau_i$.  Therefore, if the true
plaintext and corrupting vectors have the highest magnetizations
(decryption success), the Boltzmann factor $\exp[-\beta
H(\bsigma,\btau)]$ will dominate the sum over states in $Z_2$ in
the limit $\beta\to\infty$ and $\Psi(\Gamma)=0$. Alternatively, if
some other vectors $\bsigma\neq \bxi$ and $\btau\neq\bzeta$ have
the highest magnetizations of all candidates (decoding failure),
its Boltzmann factor will dominate both $Z_1$ and $Z_2$ so that
$\Psi(\Gamma)=1$. Separate temperatures $\beta_{1,2}$ and powers
$\lambda_{1,2}$ have been introduced to determine whether obtained
solutions are physical or not (values of these parameters will be
obtained via the zero-entropy condition).

To derive the average error exponent $E(p)$ we take the logarithm
of the above indicator function averaged with respect to the
disorder  variables $\Gamma=\{\bxi,\bzeta,\mcA\}$
\begin{equation}
E(p)=\lim_{M\to\infty}\frac{1}{M}\log\left\bra \Psi(\Gamma)\right\ket_{\Gamma}
\label{eq:error-exponent}
\end{equation}

The evaluation of (\ref{eq:error-exponent}) is similar in spirit
to the analysis of section \ref{sec:analysis}. For details of this
calculation we refer  the reader to~\cite{reliability} where we
also study and compare the reliability and average error exponents
of various low-density parity-check codes.

Results describing $E(p)$ for authorised decryption of the
cryptosystem \cite{KMS} are presented in figure
\ref{fig:reliability} where we plot $E(p)$ as function of the
corruption level $p$ for $(K,C,L)=(2,8,2)$ (code-rate 1/4) and
$(K,C,L)=(2,4,2)$ (code-rate 1/2) cryptosystems. It is clear that
decryption errors decay very fast with the system size as we go
away from the critical corruption level. For instance, in the case
of $R=1/4$, using a corruption level of $p=0.13$ (Shannon's limit
is at $p=0.20$) and a modest ciphertext size of $M=1000$ will
result in a negligible block error probability $P_B=10^{-11}$.

\begin{center}
\begin{figure}[t]
\setlength{\unitlength}{1.1mm}
\begin{picture}(120,55)
\put( 40,  0){\epsfysize=45\unitlength\epsfbox{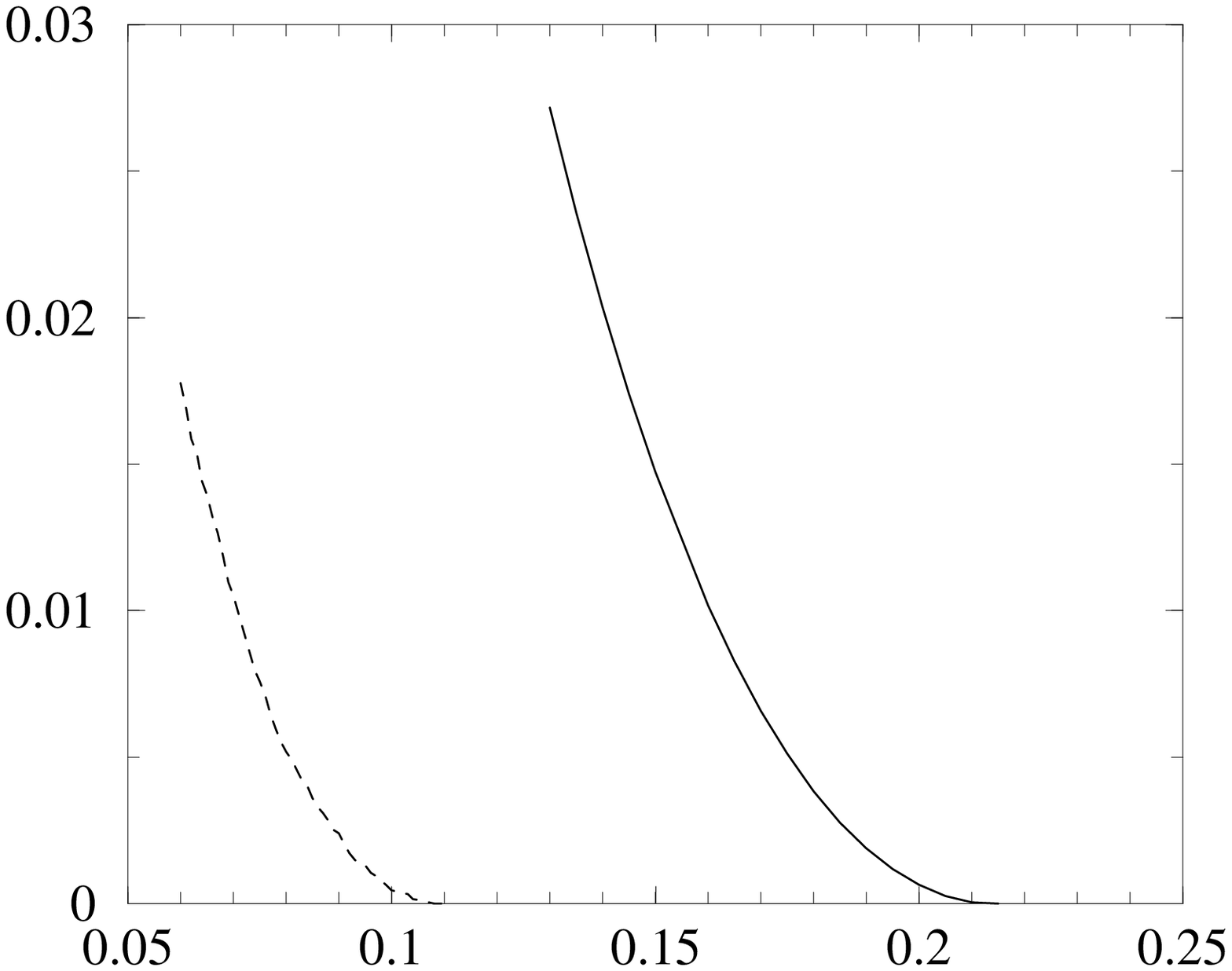}}
\put(30,27){$E(p)$}
\put(70,-2) {$p$}
\end{picture}
\caption{Reliability exponent (\ref{eq:error-exponent})
as a function of the corruption level $p$
for the case $K=L=2$ and rates $R=1/2$ (dashed line) and
$R=1/4$ (solid line). }
\label{fig:reliability}
\end{figure}
\end{center}

%--------------------------------------------------------------------%
\section{Discussion}\label{sec:conclusion}
%--------------------------------------------------------------------%

In this paper we have analyzed several security issues related to
the recently suggested public-key cryptosystem of~\cite{KMS}.
The suggested cryptosystem is based on the computational
difficulty of decomposing a dense matrix into a combination of
dense and sparse matrices (obeying certain statistics) which is a
known hard computational problem. We have considered several
attack scenarios in which unauthorized parties have acquired
partial knowledge of one or more of the private keys and/or
microscopic knowledge of the plaintext and/or the `corrupting
vector'. The analysis follows standard statistical mechanical
methods of dealing with diluted spin systems within replica
symmetric considerations. Of central importance to the
unauthorized decryption is the dynamical transition which defines
decryption success in practical situations. Our phase diagrams
show the dynamical threshold as a function of the partial acquired
knowledge of the private key; they describe regions with perfect-
($m=1$) or partial/null decryption success ($|m|<1$).

Public-key cryptosystems play an important role in modern
communications. The increasing demand for secure transmission of
information has lead to the invention of novel cryptosystems in
recent years. To this extent and based on the insight gained by
statistical physics analyses of error-correcting codes a new
family of cryptosystems was suggested in \cite{KMS}. This paper
constitutes a first step in studying  this class of cryptosystems
by considering the potential success of possible attacks.

Several future research directions aimed at improving the security
and reliability of this cryptosystem may include studying the
efficacy of irregular code constructions and the use of novel
decryption methods such as survey propagation~\cite{MPZ} for pushing
the dynamical transition point closer to the information theoretic
limits.

%--------------------------------------------------------------------%
\subsection*{Acknowledgements}
%--------------------------------------------------------------------%

We would like to thank Jort van Mourik for helpful discussions.
Support from EPSRC research grant GR/N63178, the Royal Society
(DS, NS) and Grant-in-Aid, MEXT, Japan, No. 14084206 (YK) are
gratefully acknowledged.
%DS - change 7/5/03
NS would also like to acknowledge support from the Fund for Scientific
Research-Flanders,
Belgium, for the final stages of this research.

%%%%%%%%%%%%%%%%%%%%%%%%%%%%%%%%%%%%%%%%%%%%%%%%%%%%%%%%%%%%%%%%%%%%%%%%%%%%%%%


\begin{thebibliography}{99}
\bibitem{footnote} Although larger $C$ values would increase the
absolute values of $m_{\mu i}$ and $m_{\mu j}$ in
Eq. (\ref{eq:vertical2}), this effect is relatively small and the
critical condition is determined mainly by $K$ in
Eq.~(\ref{eq:horizontal2}).
\end{thebibliography}
\end{document}